\documentclass[aps,reprint,superscriptaddress]{revtex4-1}
\usepackage{blindtext}

    \usepackage{amsmath}
    \usepackage{makeidx}
    \usepackage{amsfonts}
    \usepackage[colorlinks,hyperindex]{hyperref}
    \hypersetup
    {
        colorlinks,%
        citecolor=black,%
        linkcolor=black,%
        urlcolor=black,%
    }

\usepackage{color}
\usepackage{graphicx}
\usepackage{subfigure}

\usepackage{dcolumn}
\usepackage{amsmath}    
\usepackage{verbatim}   
\usepackage{color}      
\usepackage{hyperref}   
\usepackage{amsfonts}
\usepackage{bm}
\usepackage{titlesec}
\usepackage{capt-of}

\begin{document}


\title{Electrically driven and electrically tunable quantum light sources}

\author{J. P. Lee}\,
\affiliation{Toshiba Research Europe Limited, Cambridge Research Laboratory,\\
208 Science Park, Milton Road, Cambridge, CB4 0GZ, U.K.}
\affiliation{Engineering Department, Cambridge University,\\
9 J. J. Thomson Avenue, Cambridge, CB3 0FA, U.K.}

\author{E. Murray}\,
\affiliation{Toshiba Research Europe Limited, Cambridge Research Laboratory,\\
208 Science Park, Milton Road, Cambridge, CB4 0GZ, U.K.}
\affiliation{Cavendish Laboratory, Cambridge University,\\
J. J. Thomson Avenue, Cambridge, CB3 0HE, U.K.}

\author{A. J. Bennett}
\email{anthony.bennett@crl.toshiba.co.uk}
\affiliation{Toshiba Research Europe Limited, Cambridge Research Laboratory,\\
208 Science Park, Milton Road, Cambridge, CB4 0GZ, U.K.}

\author{D. J. P. Ellis}
\affiliation{Toshiba Research Europe Limited, Cambridge Research Laboratory,\\
208 Science Park, Milton Road, Cambridge, CB4 0GZ, U.K.}

\author{C. Dangel}
\affiliation{Toshiba Research Europe Limited, Cambridge Research Laboratory,\\
208 Science Park, Milton Road, Cambridge, CB4 0GZ, U.K.}
\affiliation{Cavendish Laboratory, Cambridge University,\\
J. J. Thomson Avenue, Cambridge, CB3 0HE, U.K.}
\affiliation{Physik Department, Technische Universit{\"a}t M{\"u}nchen, \\
85748 Garching, Germany}

\author{I. Farrer}
\thanks{Current affiliation: Department of Electronic \& Electrical Engineering, University of Sheffield, Mappin Street, Sheffield, S1 3JD, U.K. }
\affiliation{Cavendish Laboratory, Cambridge University,\\
J. J. Thomson Avenue, Cambridge, CB3 0HE, U.K.}

\author{P. Spencer}
\affiliation{Cavendish Laboratory, Cambridge University,\\
J. J. Thomson Avenue, Cambridge, CB3 0HE, U.K.}

\author{D. A. Ritchie}
\affiliation{Cavendish Laboratory, Cambridge University,\\
J. J. Thomson Avenue, Cambridge, CB3 0HE, U.K.}

\author{A. J. Shields}
\affiliation{Toshiba Research Europe Limited, Cambridge Research Laboratory,\\
208 Science Park, Milton Road, Cambridge, CB4 0GZ, U.K.}

\date{\today}%


\begin{abstract}
Compact and electrically controllable on-chip sources of indistinguishable photons are desirable for the development of integrated quantum technologies.
We demonstrate that two quantum dot light emitting diodes (LEDs) in close proximity on a single chip can function as a tunable, all-electric quantum light source. Light emitted by an electrically excited driving LED is used to excite quantum dots the neighbouring diode. The wavelength of the quantum dot emission from the neighbouring driven diode is tuned via the quantum confined Stark effect. We also show that we can electrically tune the fine structure splitting.  
\end{abstract}

\maketitle 




Sources of indistinguishable single photons or entangled photon pairs have applications in quantum key distribution \cite{bennett1984quantum, PhysRevLett.67.661}, quantum teleportation \cite{PhysRevLett.70.1895}, imaging \cite{1464-4266-4-3-372} and linear optical quantum computing \cite{knill2001scheme}. 

Self assembled quantum dots (QDs) can be used in solid-state quantum light sources. In an InAs QD the decay of a bright neutral or singly charged exciton results in the emission of a single photon \cite{0034-4885-75-12-126503}. Furthermore, the biexciton cascade of a QD has been used to generate on-demand entangled photon pairs \cite{stevenson2006semiconductor}. 
These QDs are produced using well-developed semiconductor fabrication techniques, allowing them to be integrated into semiconductor devices and photonic cavities. 

An ongoing challenge of working with self assembled quantum dots is their random distribution in the size, shape and chemical composition, which leads to a random distribution in the wavelength of the emitted photons. The distribution in transition energies is over $10^{4}$ times the width of each individual line, consequently the probability of two randomly chosen quantum dots having emission lines at the same wavelength is less than $10^{-4}$ \cite{PhysRevLett.114.150502}. Any difference in wavelength between photons makes them distinguishable.
Clearly this obstacle must be overcome in order to realise the many applications of quantum light sources that make use of Hong-Ou-Mandel (HOM) interference, which requires the photons to be indistinguishable.

Several methods of tuning the emission wavelengths of quantum dots in-situ have been explored, including varying the temperature \cite{:/content/aip/journal/apl/88/13/10.1063/1.2189747}, magnetic field \cite{PhysRevLett.103.127401}, strain \cite{Seidl200614} and electric field \cite{:/content/aip/journal/apl/78/19/10.1063/1.1369148}. The latter approach is very promising as it allows for individual tuning of many closely spaced devices suitable for producing compact integrated quantum technologies. We note that although strain tuning is usually applied to a whole device, recently a device was developed that would in principle be suitable for the production of multiple strain tuned diodes \cite{trotta2012nanomembrane}. Strain tuning can also be used in tandem with electric tuning in order to allow independent control over the exciton and biexciton energies \cite{trotta2013independent}.

\begin{figure}[h!]
\includegraphics[width=8 cm]{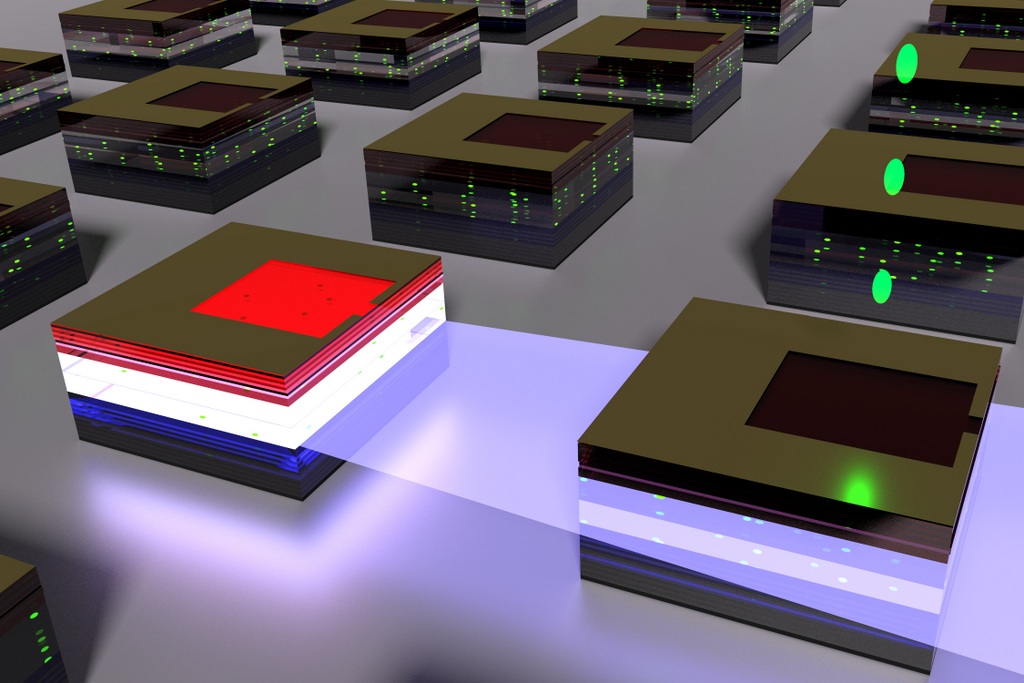}
\caption{An illustration of the devices used. The devices have a p-doped region (red), an intrinsic region (clear) and an n-doped region (blue). A LED driven strongly in forward bias (left) emits light (shown as a blue beam), which excites quantum dots in a neighbouring device (right). The quantum dots emit anti-bunched light (green).}  
\label{Fig1}
\end{figure}

\begin{figure}[h]
\includegraphics[width=8 cm]{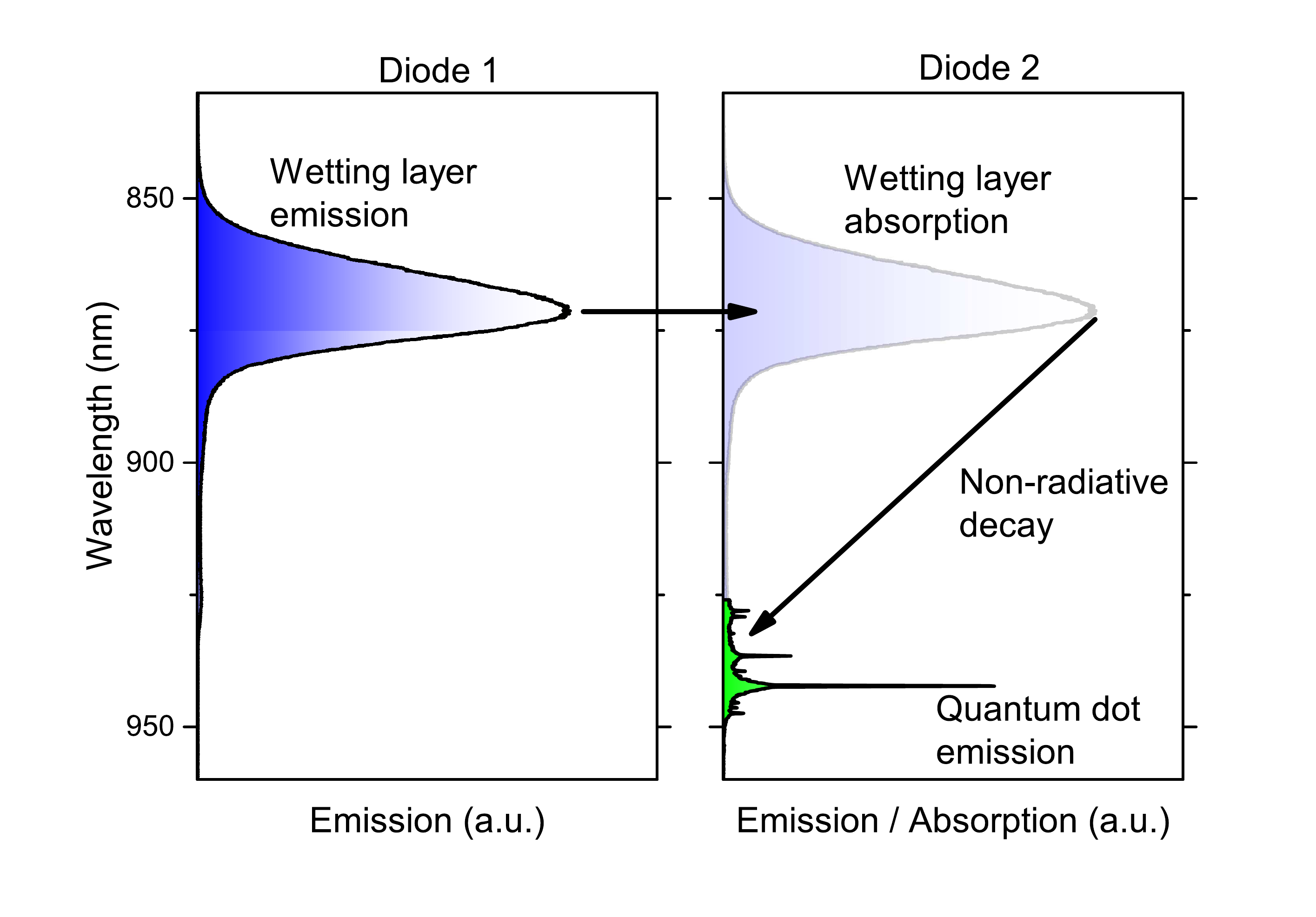}
\caption{A diagram showing the principle of operation - Light emitted from the wetting layer of Diode 1 is absorbed by the wetting layer of Diode 2, generating charge carriers in Diode 2. The charge carriers can be captured by quantum dots in Diode 2 resulting in quantum light emission. The wetting layer emission (left) and the quantum dot emission (right) are real data, the wetting layer absorption shown is a duplicate of the emission data and is included solely to show the principle of operation of the device.}  
\label{Fig2}
\end{figure}

\begin{figure}[h]
\includegraphics[width=8 cm]{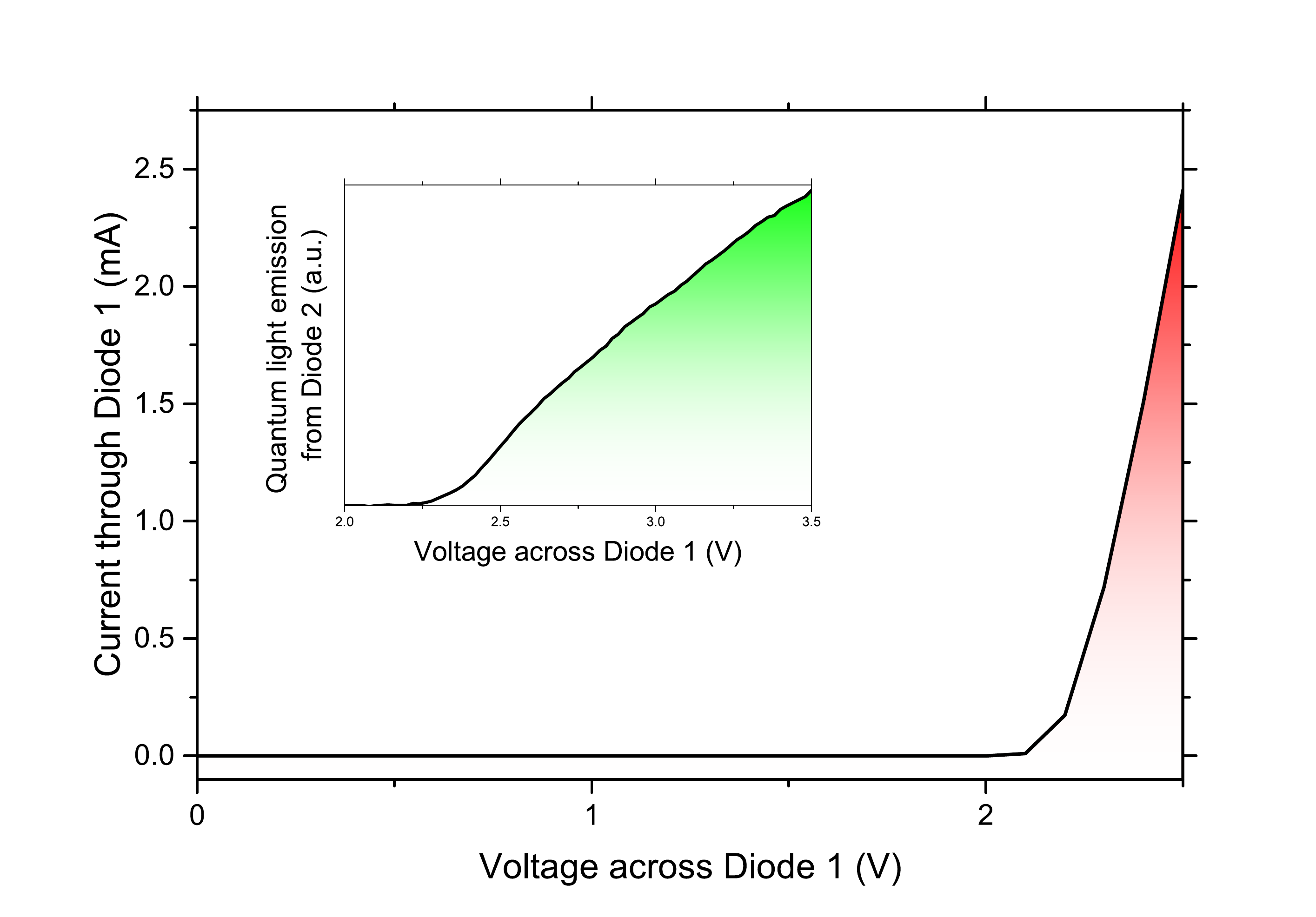}
\caption{A graph showing the current through Diode 1 as a function of voltage. The device shows diode like behaviour. {\bf Inset:} Quantum light emission from Diode 2 as a function of the voltage across Diode 1. Note that the emission and current both `turn on' at $\sim 2.25$ V.}  
\label{Fig3}
\end{figure}

\begin{figure}[h]
\includegraphics[width=8 cm]{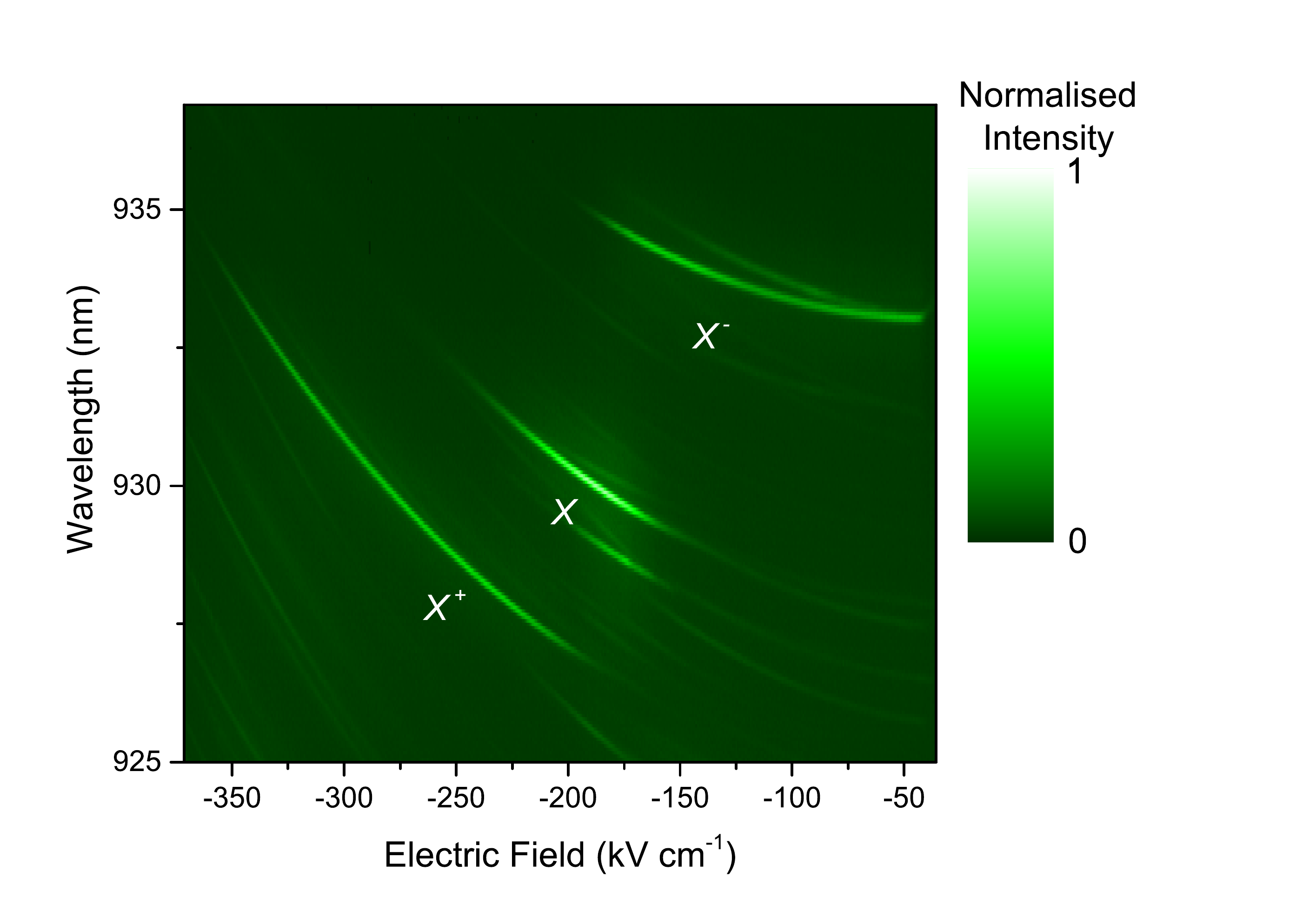}
\caption{The emission spectrum of Diode 2 (driven by Diode 1) as a function of applied electric field. Tuning of over 5 nm is visible.}  
\label{Fig5}
\end{figure}

\begin{figure}[h]
\includegraphics[width=8 cm]{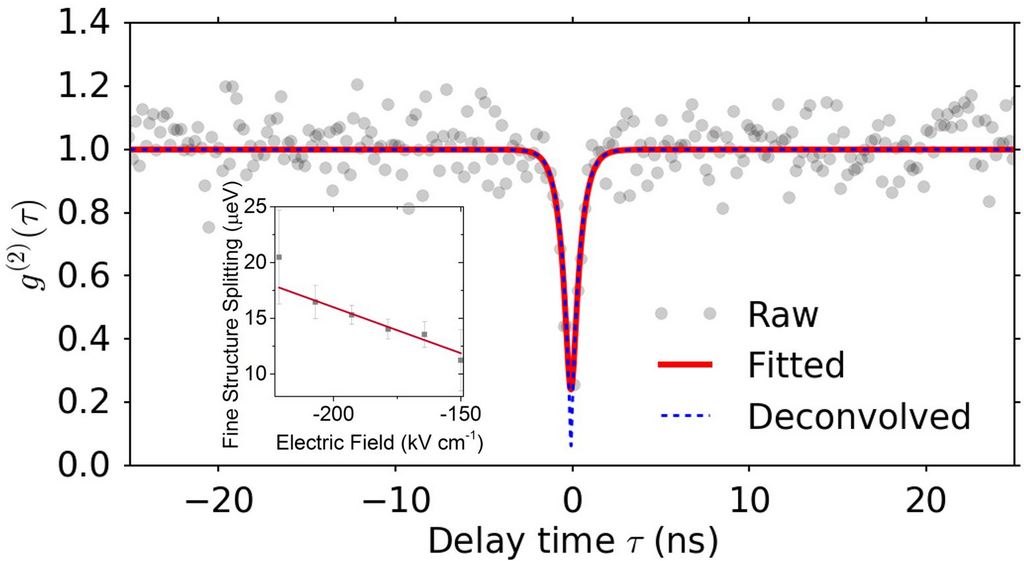}
\caption{ The second-order correlation function of a single emission line from Diode 2. {\bf Inset:)} The fine structure splitting of a neutral exciton as a function of the electric field across Diode 2.    }  
\label{Fig6}
\end{figure}

Applying a vertical electric field to a QD causes the transitions to undergo a Stark shift, which tunes the wavelength of the emitted photons. In most devices, applying an electric field of over $-60$ kV cm$^{-1}$ can result in the quenching of emission \cite{PhysRevB.66.045313}. This can be increased by an order of magnitude by embedding the quantum dots in a GaAs/AlGaAs quantum well, resulting in an increased tuning range \cite{patel2010two}. These structures have been used to demonstrate electric control of a spin \cite{de2010all}, tuning of the g-factor of trapped spins \cite{pmid:23443550} and tuning of the fine structure splitting (FSS) \cite{bennett2010electric, ward2014coherent}. A small FSS is required to produce entangled photons from the bi-exciton cascade. The key advantages of electrically tunable devices are that many individually tunable devices can be produced on a single chip using standard photo-lithographic semiconductor processing techniques and, unlike piezo-tuned systems, they do not suffer from long term drift \cite{zhang2015high}.  

For the development of compact and scalable quantum technologies, the capability to electrically trigger the generation of photons is desirable and has lead to the development of a single photon \cite{Yuan102} and entangled photon \cite{salter2010entangled} light emitting diodes. 
Combining the advantages of wavelength tuning and electrical excitation of quantum dots would allow the electrically triggered generation of indistinguishable photons from different sources.

In this paper we demonstrate a method of producing electrically triggered anti-bunched light from an electrically tunable source. The device is fabricated from a section of a single wafer and contains 16 individually tunable diode structures on a single chip.

The devices used in this study are $180\mu$m$\times 210 \mu$m planar microcavity LEDs containing a layer of InAs quantum dots embedded in a 10 nm GaAs quantum well with Al$_{0.75}$Ga$_{0.25}$As barriers. Distributed Bragg Reflectors (DBRs) grown above (4 repeats) and below the InAs (15 repeats) quantum dot layer and quantum well are used to form a half-wavelength cavity. This unbalanced cavity both increases the portion of QD light emitted vertically and acts as a horizontal waveguide for optical emission from the InAs wetting layer. The top DBR is doped p-type and the bottom DBR is doped n-type to form a diode structure suitable for electrical excitation. The key idea is to use light produced by one LED to excite the QDs in the neighbouring diode. We run one LED (Diode 1) in forward bias, resulting in broadband emission from the InAs wetting layer. The emitted light is guided horizontally due to the Bragg reflectors above and below the wetting layer. A portion of the emitted light is incident on the neighbouring LED (Diode 2) and will be absorbed by the wetting layer, generating excitons. These will be subsequently captured by quantum dots in Diode 2 resulting in quantum light emission (see figures \ref{Fig1} and \ref{Fig2}). This on-chip in plane excitation scheme is similar to the scheme presented in \cite{stock2013chip}, which makes use of a whispering-gallery-mode QD laser to excite a QD embedded in a nearby micropillar.

The cavity mode of the planar microcavity matches the emission wavelength of the quantum dots of interest, increasing the proportion of QD emission directed upwards into the collection optics. Varying the bias across Diode 2 allows wavelength tuning by Stark shifting the transitions. The quantum dots are embedded in a quantum well (as in \cite{patel2010two}) in order to increase the maximum voltage that can be applied without quenching the emission.

The intensity of light emission from Diode 2 can be controlled by varying the voltage across Diode 1 and the devices show good diode-like behaviour (figure \ref{Fig3}).
We apply a voltage of 2.5V (corresponding to a current density of $\sim 23$ A cm$^{-2}$) to Diode 1, resulting in a broad emission spectrum with a peak at around 870 nm corresponding to the InAs wetting layer (figure \ref{Fig2}). 
Figure \ref{Fig5} shows the observed emission spectrum of a QD in Diode 2 as a function of the applied electric field - a Stark shift of over 5 nm is visible.

A diffraction grating was used to spectrally filter the output of Diode 2 to pick out light from a single emission line. The filtered light was then directed into a Hanbury Brown and Twiss setup in order to perform a second-order correlation function ($g^{(2)}(\tau)$) measurement. The result can be seen in figure \ref{Fig6}). We observe a dip in $g^{(2)}(\tau)$ as $\tau$ approaches 0, giving $g^{(2)}(0) = 0.06$, below the $g^{(2)}(0) < 0.5$ threshold, which shows the output is dominated by light from a single quantum emitter. This is an order of magnitude improvement on the values of $g^{(2)}(0)$ measured in prior demonstrations of on-chip in-plane excitation \cite{munnelly2015pulsed}. We note that $g^{(2)}(0) \neq 0$, even though we have accounted for  the detector response. We attribute this to the leakage of light from Diode 1 into the output.

Finally, we demonstrate that we can tune the FSS as a function of electric field. The inset in figure \ref{Fig6} shows fine structure splitting of an exciton in Diode 2 as a function of the electric field across Diode 2, indicating that these devices could be used as sources of entangled photon pairs.

In the future we can envisage a number of changes that would increase the efficiency of the device. For example, Diode 1 could be structured to increase the directionality of emission towards Diode 2 by means of a unidirectional antenna. Similarly, a waveguide between the LEDs could increase the cross-coupling efficiency, with in principle one driving LED exciting many tunable LEDs. The use of fast electronics and low RC constant devices could allow `on-demand' emission by either varying the bias to diode 1 to modulate the `pump' or by varying the bias to Diode 2 to modulate the wavelength. 

We have demonstrated an all-electric tunable quantum source fabricated using a simple photo-lithographic process. While this is a proof-of-principle device intended to demonstrate the possibility of an electrically tunable and electrically driven quantum light source, we anticipate that future work will improve the coupling between driven and driving LEDs and lead to the production of devices suitable for compact multiplexed on-chip quantum optics technologies.

\section*{Acknowledgments}
The authors acknowledge funding from the EPSRC for MBE system used for the growth of the QD-LED.
E.M. and C.D. acknowledge support by the Marie Curie Actions within the Seventh Framework Programme for Research of the European Commission, under the Initial Training Network PICQUE (Grant No. 608062)
J. L. gratefully acknowledges financial support from the EPSRC CDT in Photonic Systems Development and Toshiba Research Europe Ltd.

\section*{Data Access}
The experimental data used to produce the figures in this paper is publicly available at \url{https://doi.org/10.17863/CAM.6837}.

\bibliographystyle{apsrev4-1}

\begin{thebibliography}{25}%
\makeatletter
\providecommand \@ifxundefined [1]{%
 \@ifx{#1\undefined}
}%
\providecommand \@ifnum [1]{%
 \ifnum #1\expandafter \@firstoftwo
 \else \expandafter \@secondoftwo
 \fi
}%
\providecommand \@ifx [1]{%
 \ifx #1\expandafter \@firstoftwo
 \else \expandafter \@secondoftwo
 \fi
}%
\providecommand \natexlab [1]{#1}%
\providecommand \enquote  [1]{``#1''}%
\providecommand \bibnamefont  [1]{#1}%
\providecommand \bibfnamefont [1]{#1}%
\providecommand \citenamefont [1]{#1}%
\providecommand \href@noop [0]{\@secondoftwo}%
\providecommand \href [0]{\begingroup \@sanitize@url \@href}%
\providecommand \@href[1]{\@@startlink{#1}\@@href}%
\providecommand \@@href[1]{\endgroup#1\@@endlink}%
\providecommand \@sanitize@url [0]{\catcode `\\12\catcode `\$12\catcode
  `\&12\catcode `\#12\catcode `\^12\catcode `\_12\catcode `\%12\relax}%
\providecommand \@@startlink[1]{}%
\providecommand \@@endlink[0]{}%
\providecommand \url  [0]{\begingroup\@sanitize@url \@url }%
\providecommand \@url [1]{\endgroup\@href {#1}{\urlprefix }}%
\providecommand \urlprefix  [0]{URL }%
\providecommand \Eprint [0]{\href }%
\providecommand \doibase [0]{http://dx.doi.org/}%
\providecommand \selectlanguage [0]{\@gobble}%
\providecommand \bibinfo  [0]{\@secondoftwo}%
\providecommand \bibfield  [0]{\@secondoftwo}%
\providecommand \translation [1]{[#1]}%
\providecommand \BibitemOpen [0]{}%
\providecommand \bibitemStop [0]{}%
\providecommand \bibitemNoStop [0]{.\EOS\space}%
\providecommand \EOS [0]{\spacefactor3000\relax}%
\providecommand \BibitemShut  [1]{\csname bibitem#1\endcsname}%
\let\auto@bib@innerbib\@empty
\bibitem [{\citenamefont {Bennett}(1984)}]{bennett1984quantum}%
  \BibitemOpen
  \bibfield  {author} {\bibinfo {author} {\bibfnamefont {C.~H.}\ \bibnamefont
  {Bennett}},\ }in\ \href@noop {} {\emph {\bibinfo {booktitle} {International
  Conference on Computer System and Signal Processing, IEEE, 1984}}}\ (\bibinfo
  {year} {1984})\ pp.\ \bibinfo {pages} {175--179}\BibitemShut {NoStop}%
\bibitem [{\citenamefont {Ekert}(1991)}]{PhysRevLett.67.661}%
  \BibitemOpen
  \bibfield  {author} {\bibinfo {author} {\bibfnamefont {A.~K.}\ \bibnamefont
  {Ekert}},\ }\href {\doibase 10.1103/PhysRevLett.67.661} {\bibfield  {journal}
  {\bibinfo  {journal} {Phys. Rev. Lett.}\ }\textbf {\bibinfo {volume} {67}},\
  \bibinfo {pages} {661} (\bibinfo {year} {1991})}\BibitemShut {NoStop}%
\bibitem [{\citenamefont {Bennett}\ \emph {et~al.}(1993)\citenamefont
  {Bennett}, \citenamefont {Brassard}, \citenamefont {Cr\'epeau}, \citenamefont
  {Jozsa}, \citenamefont {Peres},\ and\ \citenamefont
  {Wootters}}]{PhysRevLett.70.1895}%
  \BibitemOpen
  \bibfield  {author} {\bibinfo {author} {\bibfnamefont {C.~H.}\ \bibnamefont
  {Bennett}}, \bibinfo {author} {\bibfnamefont {G.}~\bibnamefont {Brassard}},
  \bibinfo {author} {\bibfnamefont {C.}~\bibnamefont {Cr\'epeau}}, \bibinfo
  {author} {\bibfnamefont {R.}~\bibnamefont {Jozsa}}, \bibinfo {author}
  {\bibfnamefont {A.}~\bibnamefont {Peres}}, \ and\ \bibinfo {author}
  {\bibfnamefont {W.~K.}\ \bibnamefont {Wootters}},\ }\href {\doibase
  10.1103/PhysRevLett.70.1895} {\bibfield  {journal} {\bibinfo  {journal}
  {Phys. Rev. Lett.}\ }\textbf {\bibinfo {volume} {70}},\ \bibinfo {pages}
  {1895} (\bibinfo {year} {1993})}\BibitemShut {NoStop}%
\bibitem [{\citenamefont {Lugiato}\ \emph {et~al.}(2002)\citenamefont
  {Lugiato}, \citenamefont {Gatti},\ and\ \citenamefont
  {Brambilla}}]{1464-4266-4-3-372}%
  \BibitemOpen
  \bibfield  {author} {\bibinfo {author} {\bibfnamefont {L.~A.}\ \bibnamefont
  {Lugiato}}, \bibinfo {author} {\bibfnamefont {A.}~\bibnamefont {Gatti}}, \
  and\ \bibinfo {author} {\bibfnamefont {E.}~\bibnamefont {Brambilla}},\ }\href
  {http://stacks.iop.org/1464-4266/4/i=3/a=372} {\bibfield  {journal} {\bibinfo
   {journal} {Journal of Optics B: Quantum and Semiclassical Optics}\ }\textbf
  {\bibinfo {volume} {4}},\ \bibinfo {pages} {S176} (\bibinfo {year}
  {2002})}\BibitemShut {NoStop}%
\bibitem [{\citenamefont {Knill}\ \emph {et~al.}(2001)\citenamefont {Knill},
  \citenamefont {Laflamme},\ and\ \citenamefont {Milburn}}]{knill2001scheme}%
  \BibitemOpen
  \bibfield  {author} {\bibinfo {author} {\bibfnamefont {E.}~\bibnamefont
  {Knill}}, \bibinfo {author} {\bibfnamefont {R.}~\bibnamefont {Laflamme}}, \
  and\ \bibinfo {author} {\bibfnamefont {G.~J.}\ \bibnamefont {Milburn}},\
  }\href@noop {} {\bibfield  {journal} {\bibinfo  {journal} {Nature}\ }\textbf
  {\bibinfo {volume} {409}},\ \bibinfo {pages} {46} (\bibinfo {year}
  {2001})}\BibitemShut {NoStop}%
\bibitem [{\citenamefont {Buckley}\ \emph {et~al.}(2012)\citenamefont
  {Buckley}, \citenamefont {Rivoire},\ and\ \citenamefont
  {Vuckovic}}]{0034-4885-75-12-126503}%
  \BibitemOpen
  \bibfield  {author} {\bibinfo {author} {\bibfnamefont {S.}~\bibnamefont
  {Buckley}}, \bibinfo {author} {\bibfnamefont {K.}~\bibnamefont {Rivoire}}, \
  and\ \bibinfo {author} {\bibfnamefont {J.}~\bibnamefont {Vuckovic}},\ }\href
  {http://stacks.iop.org/0034-4885/75/i=12/a=126503} {\bibfield  {journal}
  {\bibinfo  {journal} {Reports on Progress in Physics}\ }\textbf {\bibinfo
  {volume} {75}},\ \bibinfo {pages} {126503} (\bibinfo {year}
  {2012})}\BibitemShut {NoStop}%
\bibitem [{\citenamefont {Stevenson}\ \emph {et~al.}(2006)\citenamefont
  {Stevenson}, \citenamefont {Young}, \citenamefont {Atkinson}, \citenamefont
  {Cooper}, \citenamefont {Ritchie},\ and\ \citenamefont
  {Shields}}]{stevenson2006semiconductor}%
  \BibitemOpen
  \bibfield  {author} {\bibinfo {author} {\bibfnamefont {R.~M.}\ \bibnamefont
  {Stevenson}}, \bibinfo {author} {\bibfnamefont {R.~J.}\ \bibnamefont
  {Young}}, \bibinfo {author} {\bibfnamefont {P.}~\bibnamefont {Atkinson}},
  \bibinfo {author} {\bibfnamefont {K.}~\bibnamefont {Cooper}}, \bibinfo
  {author} {\bibfnamefont {D.~A.}\ \bibnamefont {Ritchie}}, \ and\ \bibinfo
  {author} {\bibfnamefont {A.~J.}\ \bibnamefont {Shields}},\ }\href@noop {}
  {\bibfield  {journal} {\bibinfo  {journal} {Nature}\ }\textbf {\bibinfo
  {volume} {439}},\ \bibinfo {pages} {179} (\bibinfo {year}
  {2006})}\BibitemShut {NoStop}%
\bibitem [{\citenamefont {Trotta}\ \emph {et~al.}(2015)\citenamefont {Trotta},
  \citenamefont {Mart\'{\i}n-S\'anchez}, \citenamefont {Daruka}, \citenamefont
  {Ortix},\ and\ \citenamefont {Rastelli}}]{PhysRevLett.114.150502}%
  \BibitemOpen
  \bibfield  {author} {\bibinfo {author} {\bibfnamefont {R.}~\bibnamefont
  {Trotta}}, \bibinfo {author} {\bibfnamefont {J.}~\bibnamefont
  {Mart\'{\i}n-S\'anchez}}, \bibinfo {author} {\bibfnamefont {I.}~\bibnamefont
  {Daruka}}, \bibinfo {author} {\bibfnamefont {C.}~\bibnamefont {Ortix}}, \
  and\ \bibinfo {author} {\bibfnamefont {A.}~\bibnamefont {Rastelli}},\ }\href
  {\doibase 10.1103/PhysRevLett.114.150502} {\bibfield  {journal} {\bibinfo
  {journal} {Phys. Rev. Lett.}\ }\textbf {\bibinfo {volume} {114}},\ \bibinfo
  {pages} {150502} (\bibinfo {year} {2015})}\BibitemShut {NoStop}%
\bibitem [{\citenamefont {Gevaux}\ \emph {et~al.}(2006)\citenamefont {Gevaux},
  \citenamefont {Bennett}, \citenamefont {Stevenson}, \citenamefont {Shields},
  \citenamefont {Atkinson}, \citenamefont {Griffiths}, \citenamefont
  {Anderson}, \citenamefont {Jones},\ and\ \citenamefont
  {Ritchie}}]{:/content/aip/journal/apl/88/13/10.1063/1.2189747}%
  \BibitemOpen
  \bibfield  {author} {\bibinfo {author} {\bibfnamefont {D.~G.}\ \bibnamefont
  {Gevaux}}, \bibinfo {author} {\bibfnamefont {A.~J.}\ \bibnamefont {Bennett}},
  \bibinfo {author} {\bibfnamefont {R.~M.}\ \bibnamefont {Stevenson}}, \bibinfo
  {author} {\bibfnamefont {A.~J.}\ \bibnamefont {Shields}}, \bibinfo {author}
  {\bibfnamefont {P.}~\bibnamefont {Atkinson}}, \bibinfo {author}
  {\bibfnamefont {J.}~\bibnamefont {Griffiths}}, \bibinfo {author}
  {\bibfnamefont {D.}~\bibnamefont {Anderson}}, \bibinfo {author}
  {\bibfnamefont {G.~A.~C.}\ \bibnamefont {Jones}}, \ and\ \bibinfo {author}
  {\bibfnamefont {D.~A.}\ \bibnamefont {Ritchie}},\ }\href {\doibase
  http://dx.doi.org/10.1063/1.2189747} {\bibfield  {journal} {\bibinfo
  {journal} {Applied Physics Letters}\ }\textbf {\bibinfo {volume} {88}},\
  \bibinfo {eid} {131101} (\bibinfo {year} {2006}),\
  http://dx.doi.org/10.1063/1.2189747}\BibitemShut {NoStop}%
\bibitem [{\citenamefont {Reitzenstein}\ \emph {et~al.}(2009)\citenamefont
  {Reitzenstein}, \citenamefont {M\"unch}, \citenamefont {Franeck},
  \citenamefont {Rahimi-Iman}, \citenamefont {L\"offler}, \citenamefont
  {H\"ofling}, \citenamefont {Worschech},\ and\ \citenamefont
  {Forchel}}]{PhysRevLett.103.127401}%
  \BibitemOpen
  \bibfield  {author} {\bibinfo {author} {\bibfnamefont {S.}~\bibnamefont
  {Reitzenstein}}, \bibinfo {author} {\bibfnamefont {S.}~\bibnamefont
  {M\"unch}}, \bibinfo {author} {\bibfnamefont {P.}~\bibnamefont {Franeck}},
  \bibinfo {author} {\bibfnamefont {A.}~\bibnamefont {Rahimi-Iman}}, \bibinfo
  {author} {\bibfnamefont {A.}~\bibnamefont {L\"offler}}, \bibinfo {author}
  {\bibfnamefont {S.}~\bibnamefont {H\"ofling}}, \bibinfo {author}
  {\bibfnamefont {L.}~\bibnamefont {Worschech}}, \ and\ \bibinfo {author}
  {\bibfnamefont {A.}~\bibnamefont {Forchel}},\ }\href {\doibase
  10.1103/PhysRevLett.103.127401} {\bibfield  {journal} {\bibinfo  {journal}
  {Phys. Rev. Lett.}\ }\textbf {\bibinfo {volume} {103}},\ \bibinfo {pages}
  {127401} (\bibinfo {year} {2009})}\BibitemShut {NoStop}%
\bibitem [{\citenamefont {Seidl}\ \emph {et~al.}(2006)\citenamefont {Seidl},
  \citenamefont {H{\"o}gele}, \citenamefont {Kroner}, \citenamefont {Karrai},
  \citenamefont {Badolato}, \citenamefont {Petroff},\ and\ \citenamefont
  {Warburton}}]{Seidl200614}%
  \BibitemOpen
  \bibfield  {author} {\bibinfo {author} {\bibfnamefont {S.}~\bibnamefont
  {Seidl}}, \bibinfo {author} {\bibfnamefont {A.}~\bibnamefont {H{\"o}gele}},
  \bibinfo {author} {\bibfnamefont {M.}~\bibnamefont {Kroner}}, \bibinfo
  {author} {\bibfnamefont {K.}~\bibnamefont {Karrai}}, \bibinfo {author}
  {\bibfnamefont {A.}~\bibnamefont {Badolato}}, \bibinfo {author}
  {\bibfnamefont {P.}~\bibnamefont {Petroff}}, \ and\ \bibinfo {author}
  {\bibfnamefont {R.}~\bibnamefont {Warburton}},\ }\href {\doibase
  http://dx.doi.org/10.1016/j.physe.2005.12.069} {\bibfield  {journal}
  {\bibinfo  {journal} {Physica E: Low-dimensional Systems and Nanostructures}\
  }\textbf {\bibinfo {volume} {32}},\ \bibinfo {pages} {14 } (\bibinfo {year}
  {2006})},\ \bibinfo {note} {proceedings of the 12th International Conference
  on Modulated Semiconductor StructuresProceedings of the 12th International
  Conference on Modulated Semiconductor Structures}\BibitemShut {NoStop}%
\bibitem [{\citenamefont {Findeis}\ \emph {et~al.}(2001)\citenamefont
  {Findeis}, \citenamefont {Baier}, \citenamefont {Beham}, \citenamefont
  {Zrenner},\ and\ \citenamefont
  {Abstreiter}}]{:/content/aip/journal/apl/78/19/10.1063/1.1369148}%
  \BibitemOpen
  \bibfield  {author} {\bibinfo {author} {\bibfnamefont {F.}~\bibnamefont
  {Findeis}}, \bibinfo {author} {\bibfnamefont {M.}~\bibnamefont {Baier}},
  \bibinfo {author} {\bibfnamefont {E.}~\bibnamefont {Beham}}, \bibinfo
  {author} {\bibfnamefont {A.}~\bibnamefont {Zrenner}}, \ and\ \bibinfo
  {author} {\bibfnamefont {G.}~\bibnamefont {Abstreiter}},\ }\href {\doibase
  http://dx.doi.org/10.1063/1.1369148} {\bibfield  {journal} {\bibinfo
  {journal} {Applied Physics Letters}\ }\textbf {\bibinfo {volume} {78}},\
  \bibinfo {pages} {2958} (\bibinfo {year} {2001})}\BibitemShut {NoStop}%
\bibitem [{\citenamefont {Trotta}\ \emph {et~al.}(2012)\citenamefont {Trotta},
  \citenamefont {Atkinson}, \citenamefont {Plumhof}, \citenamefont {Zallo},
  \citenamefont {Rezaev}, \citenamefont {Kumar}, \citenamefont {Baunack},
  \citenamefont {Schr{\"o}ter}, \citenamefont {Rastelli},\ and\ \citenamefont
  {Schmidt}}]{trotta2012nanomembrane}%
  \BibitemOpen
  \bibfield  {author} {\bibinfo {author} {\bibfnamefont {R.}~\bibnamefont
  {Trotta}}, \bibinfo {author} {\bibfnamefont {P.}~\bibnamefont {Atkinson}},
  \bibinfo {author} {\bibfnamefont {J.}~\bibnamefont {Plumhof}}, \bibinfo
  {author} {\bibfnamefont {E.}~\bibnamefont {Zallo}}, \bibinfo {author}
  {\bibfnamefont {R.~O.}\ \bibnamefont {Rezaev}}, \bibinfo {author}
  {\bibfnamefont {S.}~\bibnamefont {Kumar}}, \bibinfo {author} {\bibfnamefont
  {S.}~\bibnamefont {Baunack}}, \bibinfo {author} {\bibfnamefont
  {J.}~\bibnamefont {Schr{\"o}ter}}, \bibinfo {author} {\bibfnamefont
  {A.}~\bibnamefont {Rastelli}}, \ and\ \bibinfo {author} {\bibfnamefont
  {O.}~\bibnamefont {Schmidt}},\ }\href@noop {} {\bibfield  {journal} {\bibinfo
   {journal} {Advanced materials}\ }\textbf {\bibinfo {volume} {24}},\ \bibinfo
  {pages} {2668} (\bibinfo {year} {2012})}\BibitemShut {NoStop}%
\bibitem [{\citenamefont {Trotta}\ \emph {et~al.}(2013)\citenamefont {Trotta},
  \citenamefont {Zallo}, \citenamefont {Magerl}, \citenamefont {Schmidt},\ and\
  \citenamefont {Rastelli}}]{trotta2013independent}%
  \BibitemOpen
  \bibfield  {author} {\bibinfo {author} {\bibfnamefont {R.}~\bibnamefont
  {Trotta}}, \bibinfo {author} {\bibfnamefont {E.}~\bibnamefont {Zallo}},
  \bibinfo {author} {\bibfnamefont {E.}~\bibnamefont {Magerl}}, \bibinfo
  {author} {\bibfnamefont {O.~G.}\ \bibnamefont {Schmidt}}, \ and\ \bibinfo
  {author} {\bibfnamefont {A.}~\bibnamefont {Rastelli}},\ }\href@noop {}
  {\bibfield  {journal} {\bibinfo  {journal} {Physical Review B}\ }\textbf
  {\bibinfo {volume} {88}},\ \bibinfo {pages} {155312} (\bibinfo {year}
  {2013})}\BibitemShut {NoStop}%
\bibitem [{\citenamefont {Oulton}\ \emph {et~al.}(2002)\citenamefont {Oulton},
  \citenamefont {Finley}, \citenamefont {Ashmore}, \citenamefont {Gregory},
  \citenamefont {Mowbray}, \citenamefont {Skolnick}, \citenamefont {Steer},
  \citenamefont {Liew}, \citenamefont {Migliorato},\ and\ \citenamefont
  {Cullis}}]{PhysRevB.66.045313}%
  \BibitemOpen
  \bibfield  {author} {\bibinfo {author} {\bibfnamefont {R.}~\bibnamefont
  {Oulton}}, \bibinfo {author} {\bibfnamefont {J.~J.}\ \bibnamefont {Finley}},
  \bibinfo {author} {\bibfnamefont {A.~D.}\ \bibnamefont {Ashmore}}, \bibinfo
  {author} {\bibfnamefont {I.~S.}\ \bibnamefont {Gregory}}, \bibinfo {author}
  {\bibfnamefont {D.~J.}\ \bibnamefont {Mowbray}}, \bibinfo {author}
  {\bibfnamefont {M.~S.}\ \bibnamefont {Skolnick}}, \bibinfo {author}
  {\bibfnamefont {M.~J.}\ \bibnamefont {Steer}}, \bibinfo {author}
  {\bibfnamefont {S.-L.}\ \bibnamefont {Liew}}, \bibinfo {author}
  {\bibfnamefont {M.~A.}\ \bibnamefont {Migliorato}}, \ and\ \bibinfo {author}
  {\bibfnamefont {A.~J.}\ \bibnamefont {Cullis}},\ }\href {\doibase
  10.1103/PhysRevB.66.045313} {\bibfield  {journal} {\bibinfo  {journal} {Phys.
  Rev. B}\ }\textbf {\bibinfo {volume} {66}},\ \bibinfo {pages} {045313}
  (\bibinfo {year} {2002})}\BibitemShut {NoStop}%
\bibitem [{\citenamefont {Patel}\ \emph {et~al.}(2010)\citenamefont {Patel},
  \citenamefont {Bennett}, \citenamefont {Farrer}, \citenamefont {Nicoll},
  \citenamefont {Ritchie},\ and\ \citenamefont {Shields}}]{patel2010two}%
  \BibitemOpen
  \bibfield  {author} {\bibinfo {author} {\bibfnamefont {R.~B.}\ \bibnamefont
  {Patel}}, \bibinfo {author} {\bibfnamefont {A.~J.}\ \bibnamefont {Bennett}},
  \bibinfo {author} {\bibfnamefont {I.}~\bibnamefont {Farrer}}, \bibinfo
  {author} {\bibfnamefont {C.~A.}\ \bibnamefont {Nicoll}}, \bibinfo {author}
  {\bibfnamefont {D.~A.}\ \bibnamefont {Ritchie}}, \ and\ \bibinfo {author}
  {\bibfnamefont {A.~J.}\ \bibnamefont {Shields}},\ }\href@noop {} {\bibfield
  {journal} {\bibinfo  {journal} {Nature photonics}\ }\textbf {\bibinfo
  {volume} {4}},\ \bibinfo {pages} {632} (\bibinfo {year} {2010})}\BibitemShut
  {NoStop}%
\bibitem [{\citenamefont {De~La~Giroday}\ \emph {et~al.}(2010)\citenamefont
  {De~La~Giroday}, \citenamefont {Bennett}, \citenamefont {Pooley},
  \citenamefont {Stevenson}, \citenamefont {Sk{\"o}ld}, \citenamefont {Patel},
  \citenamefont {Farrer}, \citenamefont {Ritchie},\ and\ \citenamefont
  {Shields}}]{de2010all}%
  \BibitemOpen
  \bibfield  {author} {\bibinfo {author} {\bibfnamefont {A.~B.}\ \bibnamefont
  {De~La~Giroday}}, \bibinfo {author} {\bibfnamefont {A.}~\bibnamefont
  {Bennett}}, \bibinfo {author} {\bibfnamefont {M.}~\bibnamefont {Pooley}},
  \bibinfo {author} {\bibfnamefont {R.}~\bibnamefont {Stevenson}}, \bibinfo
  {author} {\bibfnamefont {N.}~\bibnamefont {Sk{\"o}ld}}, \bibinfo {author}
  {\bibfnamefont {R.}~\bibnamefont {Patel}}, \bibinfo {author} {\bibfnamefont
  {I.}~\bibnamefont {Farrer}}, \bibinfo {author} {\bibfnamefont
  {D.}~\bibnamefont {Ritchie}}, \ and\ \bibinfo {author} {\bibfnamefont
  {A.}~\bibnamefont {Shields}},\ }\href@noop {} {\bibfield  {journal} {\bibinfo
   {journal} {Physical Review B}\ }\textbf {\bibinfo {volume} {82}},\ \bibinfo
  {pages} {241301} (\bibinfo {year} {2010})}\BibitemShut {NoStop}%
\bibitem [{\citenamefont {Bennett}\ \emph {et~al.}(2013)\citenamefont
  {Bennett}, \citenamefont {Pooley}, \citenamefont {Cao}, \citenamefont
  {Sk{\"o}ld}, \citenamefont {Farrer}, \citenamefont {Ritchie},\ and\
  \citenamefont {Shields}}]{pmid:23443550}%
  \BibitemOpen
  \bibfield  {author} {\bibinfo {author} {\bibfnamefont {A.~J.}\ \bibnamefont
  {Bennett}}, \bibinfo {author} {\bibfnamefont {M.~A.}\ \bibnamefont {Pooley}},
  \bibinfo {author} {\bibfnamefont {Y.}~\bibnamefont {Cao}}, \bibinfo {author}
  {\bibfnamefont {N.}~\bibnamefont {Sk{\"o}ld}}, \bibinfo {author}
  {\bibfnamefont {I.}~\bibnamefont {Farrer}}, \bibinfo {author} {\bibfnamefont
  {D.~A.}\ \bibnamefont {Ritchie}}, \ and\ \bibinfo {author} {\bibfnamefont
  {A.~J.}\ \bibnamefont {Shields}},\ }\href {\doibase 10.1038/ncomms2519}
  {\bibfield  {journal} {\bibinfo  {journal} {Nature communications}\ }\textbf
  {\bibinfo {volume} {4}},\ \bibinfo {pages} {1522} (\bibinfo {year}
  {2013})}\BibitemShut {NoStop}%
\bibitem [{\citenamefont {Bennett}\ \emph {et~al.}(2010)\citenamefont
  {Bennett}, \citenamefont {Pooley}, \citenamefont {Stevenson}, \citenamefont
  {Ward}, \citenamefont {Patel}, \citenamefont {de~La~Giroday}, \citenamefont
  {Sk{\"o}ld}, \citenamefont {Farrer}, \citenamefont {Nicoll}, \citenamefont
  {Ritchie} \emph {et~al.}}]{bennett2010electric}%
  \BibitemOpen
  \bibfield  {author} {\bibinfo {author} {\bibfnamefont {A.}~\bibnamefont
  {Bennett}}, \bibinfo {author} {\bibfnamefont {M.}~\bibnamefont {Pooley}},
  \bibinfo {author} {\bibfnamefont {R.}~\bibnamefont {Stevenson}}, \bibinfo
  {author} {\bibfnamefont {M.}~\bibnamefont {Ward}}, \bibinfo {author}
  {\bibfnamefont {R.}~\bibnamefont {Patel}}, \bibinfo {author} {\bibfnamefont
  {A.~B.}\ \bibnamefont {de~La~Giroday}}, \bibinfo {author} {\bibfnamefont
  {N.}~\bibnamefont {Sk{\"o}ld}}, \bibinfo {author} {\bibfnamefont
  {I.}~\bibnamefont {Farrer}}, \bibinfo {author} {\bibfnamefont
  {C.}~\bibnamefont {Nicoll}}, \bibinfo {author} {\bibfnamefont
  {D.}~\bibnamefont {Ritchie}},  \emph {et~al.},\ }\href@noop {} {\bibfield
  {journal} {\bibinfo  {journal} {Nature Physics}\ }\textbf {\bibinfo {volume}
  {6}},\ \bibinfo {pages} {947} (\bibinfo {year} {2010})}\BibitemShut {NoStop}%
\bibitem [{\citenamefont {Ward}\ \emph {et~al.}(2014)\citenamefont {Ward},
  \citenamefont {Dean}, \citenamefont {Stevenson}, \citenamefont {Bennett},
  \citenamefont {Ellis}, \citenamefont {Cooper}, \citenamefont {Farrer},
  \citenamefont {Nicoll}, \citenamefont {Ritchie},\ and\ \citenamefont
  {Shields}}]{ward2014coherent}%
  \BibitemOpen
  \bibfield  {author} {\bibinfo {author} {\bibfnamefont {M.}~\bibnamefont
  {Ward}}, \bibinfo {author} {\bibfnamefont {M.}~\bibnamefont {Dean}}, \bibinfo
  {author} {\bibfnamefont {R.}~\bibnamefont {Stevenson}}, \bibinfo {author}
  {\bibfnamefont {A.}~\bibnamefont {Bennett}}, \bibinfo {author} {\bibfnamefont
  {D.}~\bibnamefont {Ellis}}, \bibinfo {author} {\bibfnamefont
  {K.}~\bibnamefont {Cooper}}, \bibinfo {author} {\bibfnamefont
  {I.}~\bibnamefont {Farrer}}, \bibinfo {author} {\bibfnamefont
  {C.}~\bibnamefont {Nicoll}}, \bibinfo {author} {\bibfnamefont
  {D.}~\bibnamefont {Ritchie}}, \ and\ \bibinfo {author} {\bibfnamefont
  {A.}~\bibnamefont {Shields}},\ }\href@noop {} {\bibfield  {journal} {\bibinfo
   {journal} {Nature communications}\ }\textbf {\bibinfo {volume} {5}}
  (\bibinfo {year} {2014})}\BibitemShut {NoStop}%
\bibitem [{\citenamefont {Zhang}\ \emph {et~al.}(2015)\citenamefont {Zhang},
  \citenamefont {Wildmann}, \citenamefont {Ding}, \citenamefont {Trotta},
  \citenamefont {Huo}, \citenamefont {Zallo}, \citenamefont {Huber},
  \citenamefont {Rastelli},\ and\ \citenamefont {Schmidt}}]{zhang2015high}%
  \BibitemOpen
  \bibfield  {author} {\bibinfo {author} {\bibfnamefont {J.}~\bibnamefont
  {Zhang}}, \bibinfo {author} {\bibfnamefont {J.~S.}\ \bibnamefont {Wildmann}},
  \bibinfo {author} {\bibfnamefont {F.}~\bibnamefont {Ding}}, \bibinfo {author}
  {\bibfnamefont {R.}~\bibnamefont {Trotta}}, \bibinfo {author} {\bibfnamefont
  {Y.}~\bibnamefont {Huo}}, \bibinfo {author} {\bibfnamefont {E.}~\bibnamefont
  {Zallo}}, \bibinfo {author} {\bibfnamefont {D.}~\bibnamefont {Huber}},
  \bibinfo {author} {\bibfnamefont {A.}~\bibnamefont {Rastelli}}, \ and\
  \bibinfo {author} {\bibfnamefont {O.~G.}\ \bibnamefont {Schmidt}},\
  }\href@noop {} {\bibfield  {journal} {\bibinfo  {journal} {Nature
  communications}\ }\textbf {\bibinfo {volume} {6}} (\bibinfo {year}
  {2015})}\BibitemShut {NoStop}%
\bibitem [{\citenamefont {Yuan}\ \emph {et~al.}(2002)\citenamefont {Yuan},
  \citenamefont {Kardynal}, \citenamefont {Stevenson}, \citenamefont {Shields},
  \citenamefont {Lobo}, \citenamefont {Cooper}, \citenamefont {Beattie},
  \citenamefont {Ritchie},\ and\ \citenamefont {Pepper}}]{Yuan102}%
  \BibitemOpen
  \bibfield  {author} {\bibinfo {author} {\bibfnamefont {Z.}~\bibnamefont
  {Yuan}}, \bibinfo {author} {\bibfnamefont {B.~E.}\ \bibnamefont {Kardynal}},
  \bibinfo {author} {\bibfnamefont {R.~M.}\ \bibnamefont {Stevenson}}, \bibinfo
  {author} {\bibfnamefont {A.~J.}\ \bibnamefont {Shields}}, \bibinfo {author}
  {\bibfnamefont {C.~J.}\ \bibnamefont {Lobo}}, \bibinfo {author}
  {\bibfnamefont {K.}~\bibnamefont {Cooper}}, \bibinfo {author} {\bibfnamefont
  {N.~S.}\ \bibnamefont {Beattie}}, \bibinfo {author} {\bibfnamefont {D.~A.}\
  \bibnamefont {Ritchie}}, \ and\ \bibinfo {author} {\bibfnamefont
  {M.}~\bibnamefont {Pepper}},\ }\href {\doibase 10.1126/science.1066790}
  {\bibfield  {journal} {\bibinfo  {journal} {Science}\ }\textbf {\bibinfo
  {volume} {295}},\ \bibinfo {pages} {102} (\bibinfo {year} {2002})},\ \Eprint
  {http://arxiv.org/abs/http://science.sciencemag.org/content/295/5552/102.full.pdf}
  {http://science.sciencemag.org/content/295/5552/102.full.pdf} \BibitemShut
  {NoStop}%
\bibitem [{\citenamefont {Salter}\ \emph {et~al.}(2010)\citenamefont {Salter},
  \citenamefont {Stevenson}, \citenamefont {Farrer}, \citenamefont {Nicoll},
  \citenamefont {Ritchie},\ and\ \citenamefont
  {Shields}}]{salter2010entangled}%
  \BibitemOpen
  \bibfield  {author} {\bibinfo {author} {\bibfnamefont {C.}~\bibnamefont
  {Salter}}, \bibinfo {author} {\bibfnamefont {R.}~\bibnamefont {Stevenson}},
  \bibinfo {author} {\bibfnamefont {I.}~\bibnamefont {Farrer}}, \bibinfo
  {author} {\bibfnamefont {C.}~\bibnamefont {Nicoll}}, \bibinfo {author}
  {\bibfnamefont {D.}~\bibnamefont {Ritchie}}, \ and\ \bibinfo {author}
  {\bibfnamefont {A.}~\bibnamefont {Shields}},\ }\href@noop {} {\bibfield
  {journal} {\bibinfo  {journal} {Nature}\ }\textbf {\bibinfo {volume} {465}},\
  \bibinfo {pages} {594} (\bibinfo {year} {2010})}\BibitemShut {NoStop}%
\bibitem [{\citenamefont {Stock}\ \emph {et~al.}(2013)\citenamefont {Stock},
  \citenamefont {Albert}, \citenamefont {Hopfmann}, \citenamefont {Lermer},
  \citenamefont {Schneider}, \citenamefont {H{\"o}fling}, \citenamefont
  {Forchel}, \citenamefont {Kamp},\ and\ \citenamefont
  {Reitzenstein}}]{stock2013chip}%
  \BibitemOpen
  \bibfield  {author} {\bibinfo {author} {\bibfnamefont {E.}~\bibnamefont
  {Stock}}, \bibinfo {author} {\bibfnamefont {F.}~\bibnamefont {Albert}},
  \bibinfo {author} {\bibfnamefont {C.}~\bibnamefont {Hopfmann}}, \bibinfo
  {author} {\bibfnamefont {M.}~\bibnamefont {Lermer}}, \bibinfo {author}
  {\bibfnamefont {C.}~\bibnamefont {Schneider}}, \bibinfo {author}
  {\bibfnamefont {S.}~\bibnamefont {H{\"o}fling}}, \bibinfo {author}
  {\bibfnamefont {A.}~\bibnamefont {Forchel}}, \bibinfo {author} {\bibfnamefont
  {M.}~\bibnamefont {Kamp}}, \ and\ \bibinfo {author} {\bibfnamefont
  {S.}~\bibnamefont {Reitzenstein}},\ }\href@noop {} {\bibfield  {journal}
  {\bibinfo  {journal} {Advanced Materials}\ }\textbf {\bibinfo {volume}
  {25}},\ \bibinfo {pages} {707} (\bibinfo {year} {2013})}\BibitemShut
  {NoStop}%
\bibitem [{\citenamefont {Munnelly}\ \emph {et~al.}(2015)\citenamefont
  {Munnelly}, \citenamefont {Heindel}, \citenamefont {Karow}, \citenamefont
  {H{\"o}fling}, \citenamefont {Kamp}, \citenamefont {Schneider},\ and\
  \citenamefont {Reitzenstein}}]{munnelly2015pulsed}%
  \BibitemOpen
  \bibfield  {author} {\bibinfo {author} {\bibfnamefont {P.}~\bibnamefont
  {Munnelly}}, \bibinfo {author} {\bibfnamefont {T.}~\bibnamefont {Heindel}},
  \bibinfo {author} {\bibfnamefont {M.~M.}\ \bibnamefont {Karow}}, \bibinfo
  {author} {\bibfnamefont {S.}~\bibnamefont {H{\"o}fling}}, \bibinfo {author}
  {\bibfnamefont {M.}~\bibnamefont {Kamp}}, \bibinfo {author} {\bibfnamefont
  {C.}~\bibnamefont {Schneider}}, \ and\ \bibinfo {author} {\bibfnamefont
  {S.}~\bibnamefont {Reitzenstein}},\ }\href@noop {} {\bibfield  {journal}
  {\bibinfo  {journal} {IEEE Journal of Selected Topics in Quantum
  Electronics}\ }\textbf {\bibinfo {volume} {21}},\ \bibinfo {pages} {681}
  (\bibinfo {year} {2015})}\BibitemShut {NoStop}%
\end{thebibliography}

%

\end{document}